\input harvmac.tex

\def\goto{\rightarrow}

\noblackbox

\Title{PUPT-1739, hep-th/9711046}
{\vbox{\centerline{Finite-Time Amplitudes In Matrix Theory}}}
\centerline{\O yvind Tafjord and Vipul Periwal
\footnote{$^\star$}
{otafjord@princeton.edu, vipul@phoenix.princeton.edu,
}}
\bigskip\centerline{Department of Physics}
\centerline{Princeton University}
\centerline{Princeton, NJ 08544}

\vskip .3in
\noindent
We evaluate one-loop finite-time amplitudes for graviton scattering in
Matrix theory and compare to the corresponding amplitudes in
supergravity. We find agreement for arbitrary time intervals at
leading order in distance, providing a functional agreement
between supergravity and Matrix theory.
At subleading order, we find corrections to
the effective potential found from previous phase shift calculations
in
Matrix theory.

\Date{10/97} 

\newsec{Introduction}


\nref\bfss{T.  Banks, W.  Fischler, S.  Shenker, and L.  Susskind,
{\sl Phys. Rev.} {\bf D55} (1997) 5112, hep-th/9610043.}
\nref\finite{L. Susskind, {\it Another conjecture about M(atrix)
theory}, hep-th/9704080}
\nref\state{T. Banks, {\it The state of Matrix theory},
hep-th/9706168}
\nref\dkps{M.R. Douglas, D. Kabat, P. Pouliot and S.H. Shenker, {\sl
Nucl. Phys.} {\bf B485} (1997) 85, hep-th/9608024}
\nref\aharony{O. Aharony and M. Berkooz, {\sl Nucl. Phys.} {\bf
B491} (1997) 184, hep-th/9611215}
\nref\mathur{G. Lifschytz and S.D. Mathur, {\it
Supersymmetry and Membrane Interactions in M(atrix) Theory},
hep-th/9612087}
\nref\tseytl{I. Chepelev and A.A. Tseytlin, {\it Long-distance
Interactions of D-brane Bound States and Longitudinal 5-brane in
M(atrix) Theory}, hep-th/9704127}
\nref\krauss{P. Kraus, {\it Spin-Orbit Interaction from Matrix
Theory}, hep-th/9709199 }
\nref\polpou{J. Polchinski and P. Pouliot, {\it Membrane Scattering
with M-Momentum Transfer}, hep-th/9704029}
\nref\bb{K. Becker and M. Becker, {\it A two-loop test of M(atrix)
theory}, hep-th/9705091}
\nref\bbpt{K. Becker, M. Becker, J. Polchinski and A.A. Tseytlin,
{\sl Phys. Rev.} {\bf D56} (1997) 3174 }
\nref\seiberg{N. Seiberg, {\it Why is the Matrix model correct?},
hep-th/9710009}
\nref\alwis{S.P. de Alwis, {\it Matrix Models and String World Sheet
Duality}, hep-th/9710219}
\nref\tseytlb{I. Chepelev and A.A. Tseytlin, {\it Long-distance
interactions of branes: correspondence between supergravity and super
Yang-Mills descriptions}, hep-th/9709087}
\nref\juan{J.M. Maldacena, {\it Branes Probing Black Holes},
hep-th/9709099}

Matrix theory \bfss\ is a remarkable proposal for a non-perturbative
definition of M-theory, a supposedly consistent 11-dimensional quantum
theory having supergravity as its low energy limit. Matrix theory is
defined as the maximally supersymmetric quantum mechanics of $U(N)$
matrices, describing the lowest states of open strings connecting $N$
D0-branes. Supergravitons, for instance, appear as bound states of
these
 D0-branes. The original conjecture of Banks, Fischler, Shenker
and
Susskind \bfss\ relates the large $N$ limit
of Matrix theory to M-theory in the Infinite Momemtum Frame. Later,
Susskind \finite\ expanded the conjecture to a relation between finite
$N$ Matrix theory and M-theory with a compact null-direction.

There have been several successful tests of these conjectures so far,
see, e.g., \state\ for an overview. Part of the original evidence
involved comparing graviton scattering phase shifts computed in Matrix
theory and supergravity in the limit of large impact parameter. The
fact that these agree originates from the work of Douglas, Kabat,
Pouliot, and Shenker \dkps, where it was shown how supersymmetry leads
to a cancellation of the contribution from the massive modes of the
strings connecting the D0-branes, implying that the long distance
behavior can be reproduced by keeping only the lowest open string
modes. These phase shifts corresponds to an eikonal approximation
where the gravitons move along infinite straight lines at large impact
parameter. Similar calculations have later successfully extended this
to processes involving higher branes (as in
\refs{\aharony,\mathur,\tseytl}) and spin effects
\krauss, as well as processes involving longitudinal momentum
transfer \polpou.  The Matrix theory calculation of the graviton
scattering process has lately been pushed to two loops
\refs{\bb,\bbpt}, giving further evidence for the finite $N$
conjecture. Recently Seiberg \seiberg\ has given arguments for why the
Matrix theory conjecture is correct, and this was further examined in
the context of graviton scattering in \alwis.

\nref\aspect{S. Coleman, {\it Aspects of Symmetry}, Cambridge
University Press (Cambridge, U.K., 1985)}

{}From the infinite time phase shift calculation in Matrix theory one
derives an effective potential between gravitons of relative velocity
$v$ and relative distance $r$. This potential is a double expansion in
$v^2/r^4$ and $1/r^3$ (the latter is the loop expansion), starting out
as \bbpt
\eqn\dblexp{\eqalign{
-V_{\rm eff}&={15\over16}\kappa^2{v^4\over r^7}
+0\cdot\kappa^{8/3}{v^6\over r^{11}}
+{9009\over4096}\kappa^{10/3}{v^8\over r^{15}}
+{\cal O}(\kappa^4{v^{10}\over r^{17}})\cr
&+0\cdot\kappa^{10/3}{v^4\over r^{10}}+{225\over32}\kappa^4{v^6\over
r^{14}}
+{\cal O}(\kappa^{14/3}{v^8\over r^{17}})\cr
&+{\cal O}(\kappa^{14/3}{v^4\over r^{13}}),\cr}}
where we have indicated how each term scales with the 11-dimensional
gravitational coupling, otherwise (when setting $\kappa=1$ above) the
units are as in \bb. So far, the two first terms along the diagonal of
this double expansion, with integer powers of $\kappa$, have been
found to coincide with supergravity \bbpt. It has been argued that all
the terms along the diagonal will agree with classical supergravity
and that all terms to the left of the diagonal should vanish
\refs{\tseytlb,\juan,\alwis}, while other non-zero terms in this
expansion
should come from higher-derivative terms of the supergravity effective
actions \alwis.

These infinite time straight line phase shift calculations remain one
of very few quantities one can actually calculate directly in Matrix
theory. It is important to try to extend this repertoire as much as
possible.  In this paper we will consider a finite time version of
this calculation, and from the resulting ``phase shift'' we can read
off an effective potential locally, rather than integrated along an
infinite path. We will only work to one loop, and there we find again
the ${15\over16}{v^4\over r^7}$ potential of supergravity to leading
order, for arbitrary time intervals. This provides a stronger
equivalence between Matrix theory and supergravity than is
demonstrated from the infinite-time case, as the finite time amplitude
contains more information.  Of course, the main conjecture of \bfss\
concerns the $S$-matrix, so this result suggests that their conjecture
could perhaps be strengthened.  However, higher order terms in the
potential will be modified compared to the infinite-time case, by
interesting terms that integrate to zero along the infinite line, thus
showing that the complete matching of the leading term is
non-trivial. Working on a finite time interval means that we have to
be careful about what boundary conditions are put on the high energy
modes that are integrated out. We will consider the most natural
boundary condition for comparing to supergravity, and also investigate
how the answer varies with other choices. We find that the leading
term is robust towards changing the boundary conditions, while the
subleading terms are more sensitive.  In the large $N$ limit proposed
in \bfss, these terms are also subleading in powers of $N$,
but they may be interesting in
Susskind's finite $N$ conjecture \finite.
 Note that if the Planck scale $l_P$ is the only
relevant scale in   supergravity, then since $\kappa_{11}^{2/3}\sim
l_P^3$ these terms seem  to indicate corrections to the
supergravity action dimensionally going like $R+R^{5/2}+\ldots$ rather
than $R+R^2+\ldots$.

This paper is organized as follows: In sect.~2, we set up the basic
problem.  Sect.~3 contains our calculations, and sect.~4 has some
concluding remarks.

\nref\claudson{M. Claudson and M.B. Halpern, {\sl
Nucl. Phys.} {\bf B250} (1985) 689}

\newsec{Comparing Matrix theory and supergravity}

Graviton scattering in Matrix theory is described by $U(N_1+N_2)$
supersymmetric quantum mechanics, where $N_1/R$ and $N_2/R$
are the longitudinal lightlike momenta of the gravitons, $R$ being the
radius of the compact null-direction.
We will take $N_1=N_2=1$ here,
as usual the leading order factors of $N_i$ can be
easily reinstated afterwards. The bosonic part of the $U(2)$ action is
given by \claudson
\eqn\action{S=\Tr\int dt\left({1\over2}(D_t
X^i)^2+{1\over4}[X^i,X^j]^2\right),}
where for simplicity we suppress dependences on $R$ and the
11-dimensional Planck scale (see \bbpt). We defined $D_t
X^i=\partial_t
X^i +[A,X^i]$, with $A,\,X^i$ being $U(2)$ matrices which we can
decompose as
\eqn\decomp{X^i={i\over 2}\left(X_0^i I+X_a^i \sigma^a\right).}
The $\sigma^a,\ a=1\ldots3$, are the Pauli matrices. There are also
corresponding fermionic fields $\psi$.  The $X_0^i$ fields describe
the center of mass motion, and they (together with $A_0$ and $\psi_0$)
decouple from the rest and will be ignored from now on. We interpret
$X_3^i$ as the relative coordinate of the gravitons, while the
corresponding gauge field $A_3$ can in principle be gauged away (in
background field gauge, which we employ here, it is set to a
constant). The fermionic $\psi_3$ fields incorporate the spin and
polarization of the (super-)gravitons. The off-diagonal 1,2 modes do
not have such an intuitive description in the long distance
supergravity, in fact they can be thought of as arising from open
string
considerations valid only at short distances, and they then represent
the
unexcited modes of these open strings. When the gravitons are far
part, the off-diagonal modes are all very heavy, with masses roughly
proportional to the distance. One can then imagine integrating these
modes out in a Born-Oppenheimer type of approach, as (non-local)
internal degrees of freedom. This leaves a theory only involving the
diagonal modes that are interpreted as the positions of supergravitons. This
separation of light and heavy modes is only exact in the limit of
large ${<}X_3^i{>}$ though, and a more systematic
understanding of this is needed, especially at higher loop orders.

When integrating out the off-diagonal modes, we need to supply
boundary conditions on the fields. In previous infinite time
phase shift calculations, the fields are taken to vanish at
$\pm\infty$, but for finite time intervals we should be more
careful. The off-diagonal modes are at one loop level described by
harmonic oscillators, with frequency proportional to their mass, and
specifying boundary conditions can be done by specifying the initial
and
final states of the harmonic oscillators. These harmonic oscillator
modes do not appear at all in supergravity, so exciting them out of
their ground state would seem to take us out of the realm of
supergravity. For comparing finite time amplitudes, the most natural
assumption therefore appears to be to put these modes in their ground
state for both the initial and final states. We will also consider
other choices of boundary conditions.

If we also specify boundary conditions for $X_3^i$ and integrate it
out,
we will find the amplitude for a finite-time graviton propagation. We
can write this schematically as (we now work in Euclidean time)
\eqn\ampl{
\left< X_{3,s}^i,T_s|X_{3,f}^i,T_f\right>={\cal
N}\int_{X_3(T_s)=X_{3,s}}^{X_3(T_f)=X_{3,f}}\!\!\!\!\!\!{\cal D}X_a
e^{-S(X_a)}
\equiv{\cal N}\int_{X_{3,s}}^{X_{3,f}}{\cal D}X_3 e^{-S_{\rm
eff}(X_3)}
\equiv e^{-S_0+\delta}.}
Here $S_0$ is the action of the classical trajectory implementing the
boundary conditions, and in an abuse of language we call $\delta$ the
finite time phase shift and try to relate it to an effective potential
through $\delta=-\int V_{\rm eff}$. Since we are not considering any
sort of spin effects, $\psi_3$ (as well as $A_3$, the ghost $C_3$ and
$X_3$ once the boundary conditions have been implemented) just go
along for the ride as massless free fields here at one-loop level.

In supergravity the most convenient way to do a similar
computation, is
to give one of the gravitons a large longitudinal momentum such that
it can be treated as a classical source for the other graviton. One
can then derive an action for the probe graviton moving in the field
of this source, $S_{\rm SG}(X_3^i)$. There are then two ways to view
the comparison between Matrix theory and supergravity. On one hand, we
can directly compare the effective actions of the light modes $S_{\rm
eff}(X_3^i)$ and $S_{\rm SG}(X_3^i)$ for various paths $X_3^i(t)$ and
see how they match---our computations in this paper are equivalent to
doing this for a finite line segment. This seems to be the cleanest
way of interpreting our results. Alternatively we can evaluate the
finite time amplitude in supergravity, which, at the semiclassical
level,
is given by integrating the action along the geodesic connecting the
two space time points. At large separation of the gravitons, the
leading order amplitude can be found using an eikonal approximation,
replacing the geodesic by a constant velocity straight line.

We will
choose coordinates such that the straight line connecting the two
space time points is given by
\eqn\xthree{\eqalign{
X_{3,0}^1(t)&=vt,\cr
X_{3,0}^2(t)&=b,\cr
X_{3,0}^{i{>}2}(t)&=0.\cr}}

\nref\relbrane{V. Balasubramanian and F. Larsen, {\it Relativistic
brane scattering}, hep-th/9704143}

The supergravity action for the probe graviton is given in \bbpt, and
we find that the supergravity action evaluated for a finite time line
segment
(alternatively, an eikonal approximation to the finite time phase
shift)
is given by
\eqn\deltasg{
\delta_{\rm SG}=\int_{T_s}^{T_f}\left[{15\over16}{v^4\over r^7(t)}
+{\cal O}({v^6\over r^{14}})\right]dt,}
where $r(t)=\sqrt{v^2t^2+b^2}$. The ${\cal O}({v^6\over r^{14}})$
term we
don't expect to see until two loops in Matrix theory and will not
concern us here. Note that, as argued in \relbrane, the corrections
to the eikonal approximation are of order $v^2/r^7$ and thus fall
along the diagonal in \dblexp\ and will not be of concern to us here,
either.

\newsec{Finite-time calculation in Matrix theory}

On the Matrix theory side, to evaluate \ampl\ we implement the
boundary
conditions on $X_3^i$ by a suitable solution to the equations of
motion. These solutions are straight lines, so we expand about
$X_{3,0}^i(t)$ in \xthree. The path integral can then be evaluated as
in \dkps, at one loop it yields a product of determinants. The
phase shift is given by
\eqn\dmti{\eqalign{\delta=\ln[&\det{}\!^{-6}({-}\partial_t^2{+}r^2)
\det{}\!^{-1}({-}\partial_t^2{+}r^2{+}2v)
\det{}\!^{-1}({-}\partial_t^2{+}r^2{-}2v) \cr
\times&\det{}\!^4({-}\partial_t^2{+}r^2{+}v)
\det{}\!^4({-}\partial_t^2{+}r^2{-}v)].}}
Now these determinants represents the evolution of a time
dependent harmonic superoscillator from a state $\left|\psi_s\right>$
at
time $T_s$ to a state $\left|\psi_f\right>$ at time $T_f$. The
boundary conditions we put on the fermionic part should be determined
by supersymmetry from the boundary conditions on the bosonic part. We
here choose to represent each determinant by a separate bosonic
harmonic oscillator of the appropriate frequency, evolving from
corresponding initial states to final states. This prescription
preserves a symmetry between all the modes, and we believe it gives
equivalent results to treating the full superoscillator. We write
this as
\eqn\detdef{\eqalign{
\det{}\!^{-{1\over2}}&(-\partial_t^2+\omega^2(t))=
\left<\psi_f\right|{\rm T}\exp\left[-\int_{T_s}^{T_f}H(t)dt\right]
\left|\psi_s\right>\cr
&={\cal N}\int d\varphi_f
d\varphi_s\psi^*_f(\varphi_f)\psi_s(\varphi_s)
\int_{\varphi_s}^{\varphi_f}{\cal D}\varphi(t)
\exp\left[-\int_{T_s}^{T_f}L(t)dt\right]\cr
&={\cal N}\int d\varphi_f
d\varphi_s\psi^*_f(\varphi_f)\psi_s(\varphi_s)
\exp\left[-{1\over2}(\varphi_f\dot{\varphi}_0(T_f)-
\varphi_s\dot{\varphi}_0(T_s))\right]
\det{}\!^{-{1\over2}}_0(-\partial_t^2+\omega^2(t)),\cr}}
where
\eqn\defdef{\eqalign{
H&={1\over2}\left(p^2+\omega^2\varphi^2\right),\ \
p=-i\partial_\varphi,\cr
L&={1\over2}\left(\dot{\varphi}^2+\omega^2\varphi^2\right),\cr
\left[-\partial_t^2+\omega^2(t)\right]\varphi_0(t)&=0,\ \
\varphi_0(T_s)=\varphi_s,\ \varphi_0(T_f)=\varphi_f.\cr}}
By $\det_0$ we denote the determinant evaluated on the space of
functions that vanish at both endpoints.

In order to extract the leading order supergravity potential, we
actually need to do very little work, the simplest adiabatic
approximation is sufficient. In this approximation, we assume the
harmonic oscillator to stay in its ground state throughout, and we
find the estimate
\eqn\adiab{
\det{}\!^{-{1\over2}}(-\partial^2+\omega^2(t))\approx\exp\left[
-\int_{T_s}^{T_f}{1\over2}\omega(t)dt\right].}
Using $\omega^2(t)=v^2t^2+b^2+\alpha v$, inserting this into \dmti,
and
expanding in powers of $v$, we get the phase shift
\eqn\dmtb{
\delta_{ad}=
\int_{T_s}^{T_f}dt\left[
{15\over16}{v^4\over r^7}+{315\over128}{v^6\over r^{11}}
+{27027\over4096}{v^8\over r^{15}}+\ldots\right].}  We see that the
leading term at large distance agrees with the supergravity result
\deltasg, while the higher terms do not even agree with the
infinite time phase shift calculation.  This way of deriving the
potential energy between the gravitons is tantamount to summing up the
zero point energy of the harmonic oscillators representing the
off-diagonal modes, which is how various potential energies were
computed for instance in \aharony.  We see that at subleading order
there is a distinction between these two methods of extracting an
effective potential.

We will now embark on a more systematic evaluation of the finite time
amplitude. We will use two approaches, valid, roughly speaking, at
long and short time intervals respectively.

\nref\absteg{M.\ Abramowitz and I.A.\ Stegun, {\it Handbook of
Mathematical Functions}, Dover, New York, 1965}
\nref\work{Work in progress}

For long time intervals, we use the last line of \detdef. The
classical trajectory $\varphi_0(t)$ is a parabolic cylinder function
\absteg. We can systematically solve for it using a WKB type expansion
(equivalent to the Darwin expansion in \absteg). To this end, write
\eqn\vphi{
\varphi_0(t)=\exp\left[\int_{T_s}^t\sum_{n=0}^{\infty}
\hbar^{n-1}f_n(t')dt'\right],}
and consider $\omega(t)$ to be of order $1/\hbar$. Solving the
equation
for $\varphi_0$ in \defdef\ order by order in $\hbar$ (and then
setting
$\hbar=1$), we find the first few $f$'s
\eqn\wkbsol{\eqalign{
f_0&=\omega,\cr
f_1&=-{v^2t\over2\omega^2},\cr
f_2&={v^2\over8\omega^5}(2\omega^2-5v^2t^2),\cr
f_3&={3v^4t\over8\omega^8}(3\omega^2-5v^2t^2),\cr
f_4&={v^4\over128\omega^{11}}(-76\omega^4+884\omega^2v^2t^2-1105v^4t^4).\cr}}

There is also another solution obtained by $f_{2n}\goto-f_{2n}$ for
all $n$.  We see that
\eqn\wkbsys{
f_n\sim\omega\left({v\over\omega^2}\right)^n g_n(vt/\omega),}
where $g_n$ only contains non-negative powers of its argument. Thus
this is an expansion in $v/\omega^2$ which is what we want. If we
define
\eqn\abdef{
A(t)=\sum f_{2n}(t),\ \ B(t)=\sum f_{2n+1}(t),}
the general solution can be written
\eqn\vphigen{
\varphi_0(t)=c_1e^{\int_{T_s}^t(B+A)}+c_2e^{\int_{T_s}^t(B-A)}.}
Solving the boundary conditions, we can evaluate
\eqn\vphider{\eqalign{
\dot{\varphi}_0(T_s)&=\varphi_s B_s-\varphi_s A_s\coth\int A+
{\varphi_f A_s e^{-\int B}\over\sinh\int A},\cr
\dot{\varphi}_0(T_f)&=\varphi_f B_f+\varphi_f A_f\coth\int A-
{\varphi_s A_f e^{\int B}\over\sinh\int A}.\cr}} Here $A_s\equiv
A(T_s)$
and so on.  It is now straightforward to evaluate the integral over
$\varphi_{s,f}$ in \detdef. The ground state wave function is
\eqn\grstwf{
\psi(\varphi)=\left({\omega\over\pi}\right)^{1/4}e^{-{1\over2}\omega
\varphi^2},}
and the integral is gaussian. The remaining determinant
$\det_0(-\partial_t^2+\omega^2(t))$ can be evaluated using the method
explained in \aspect, p.\ 340. The prescription is to
solve the equation
\eqn\deteq{
\left(-\partial_t^2+\omega^2(t)\right)\chi(t)=0;\ \
\chi(T_s)=0, \dot{\chi}(T_s)=1,}
then
\eqn\deteqb{
\det{}\!_0(-\partial_t^2+\omega^2(t))={\cal N'}\chi(T_f).}
We already solved this equation above. Solving for the boundary
conditions, we find
\eqn\deteqc{
\chi(T_f)={e^{\int B}\over A_s}\sinh\int
A={1\over\sqrt{A_sA_f}}\sinh\int A,}
using $B=-\dot{A}/(2A)$ which is easily derived from the equation.

We now restrict to the regime where $\int\omega\gg1$, we can then
replace all hyperbolic functions by exponentials up to errors of order
$\exp(-2\int\omega)$. Doing the gaussian integral, we find
\eqn\detres{\eqalign{
\det&(-\partial_t^2+\omega^2(t))={\cal N}''
{\left[1+{A_s'-B_s\over2\omega_s}\right]
\left[1+{A_f'+B_f\over2\omega_f}\right]\over\sqrt{\left[1+{A_s'\over\omega_s}
\right]\left[1+{A_f'\over\omega_f}\right]}
}
\exp\left[\int(\omega+A')\right][1+{\cal O}
(e^{-2\int\omega})]\cr
&={\cal N}''
\exp\left[
\int_{T_s}^{T_f}\omega
dt+{1\over8}\int_{T_s}^{T_f}{v^4t^2\over\omega^5}dt
-{v^4\over32}\left({T_f^2\over\omega_f^6}+{T_s^2\over\omega_s^6}\right)
+{\cal O}(v^3/\omega^6)+{\cal O}(e^{-2\int\omega})\right].\cr}} Here
we defined $A'=A-f_0$ and $B'=B-f_1$. The result takes a simple
symmetric form, every second term in the expansion can be written as
an integral, while the others have equal contributions from each
endpoint (the first of these actually cancels completely). This latter
type of term could in principle be removed by a suitable normalization
of the initial and final
 wave functions we use, however as far as we can see, there is
no rationale for doing that. There are no ambiguous time-dependent
normalization factors floating around here, as we see from comparison
with the short-time calculation below.
When we write ${\cal
O}(v^3/\omega^6)$, it is meant up to positive powers of
$vt/\omega$. We can now plug this into
\dmti\ and expand in powers of $v$, this gives the phase shift
\eqn\dmtc{\eqalign{
\delta&=\int_{T_s}^{T_f}{15\over16}{v^4\over r^7}dt+\int_{T_s}^{T_f}
{315\over128}
{v^6\over r^{11}}(1-11(\hat{r}\cdot\hat{v})^2)dt
+{45\over4}v^8\left({T_f^2\over r_f^{14}}+{T_s^2\over
r_s^{14}}\right)\cr
&+\int_{T_s}^{T_f}
{429\over4096}{v^8\over r^{15}}[63+1500(\hat{r}\cdot\hat{v})^2
-12070(\hat{r}\cdot\hat{v})^4]dt+{\cal O}(v^8/r^{16})
+{\cal O}(e^{-2\int r}).\cr}}
Here $\hat{r}=\vec{r}/r$. This result reduces to the first line of
\dblexp\ in the limit $T_s\goto-\infty$, $T_f\goto\infty$ as it
should, but we see the interesting feature that it differs for finite
time intervals. The leading order potential still matches supergravity
perfectly. The subleading potential is different from both the
instanteneous zero-point energy potential and the potential derived
from the infinite time phase shift---in a sense it interpolates
between these two.
There are also terms depending on the end points additively
and thus cannot be written as an integrated potential. The
interpretation of these terms is not clear.
Note that the first subleading term is a simple total
derivative,
\eqn\deriv{
{v^6\over r^{11}}(1-11(\hat{r}\cdot\hat{v})^2)={d\over dt}{v^5\over
r^{10}}\hat{r}\cdot\hat{v}}
for the straight-line trajectory.  It seems unlikely that the
corresponding term for a curved trajectory would be a total
derivative \work.

It is interesting to see how the result changes for different choices
of boundary conditions on the off-diagonal fields. If we choose
boundary conditions such that the fields vanish at the end points, the
determinants are just given by $\det_0$, and we find the phase shift
\eqn\dmtvv{
\delta_{vv}=\int_{T_s}^{T_f}{15\over16}{v^4\over r^7}dt
-3v^4\left({1\over r_f^8}+{1\over r_s^8}\right)
-\int_{T_s}^{T_f}{1575\over128}
{v^6\over r^{11}}(1-11(\hat{r}\cdot\hat{v})^2)dt+\ldots.}  This also
reduces to \dblexp\ in the infinite time limit, as expected, and the
leading potential is unchanged, but the subleading terms have
changed. We can also imagine picking boundary conditions such that we
have the ground state initially while summing over all final states,
this corresponds to putting $\psi_f(\varphi_f)=1$ in \detdef. Also in
this
case we find the same leading order potential, while the higher terms
are now  asymmetric in $T_f$ and $T_s$ and cannot be written in a
particularly illuminating form.

The approach above breaks down for $\Delta T=T_f-T_s$ very small.
We want to check further what the Matrix theory predicts in this
regime. In the infinite-momentum frame, it is not clear to us
whether physics at short time scales as described by Matrix theory
should agree with physics at short time scales as described by
supergravity.  Since supergravity is an effective low-energy
description of M-theory,  one might expect Matrix theory to behave
differently
from supergravity in this regime.
To deal with short time-intervals, we use the
first representation of the determinant in \detdef. We write
\eqn\hamsplit{\eqalign{
H&=H_s+H_p,\cr
H_s&={1\over2}(p^2+\omega_s^2\varphi^2),\cr
H_p&={1\over2}v^2(t^2-T_s^2)\varphi^2,}}
and treat $H_p$ as a time-dependent perturbation. This expansion
should be useful for $v^2(T_f^2-T_s^2)$ small, in particular
$\omega_s\Delta T$ can be allowed large thus giving an overlap in the
region of validity with
the expansion considered above. We expand the time ordered product as
\eqn\exptp{
\det{}\!^{-{1\over2}}(-\partial_t^2+\omega^2(t))=
\left<0_f\right|e^{-H_s\Delta T}\left|0_s\right>-\int_{T_s}^{T_f}
\left<0_f\right|e^{-H_s(T_f-t)}H_p(t)e^{-H_s(t-T_s)}\left|0_s\right>
+\ldots.}
Here $\left|n_s\right>$ denotes the energy eigenstates of the harmonic
oscillator at time $T_s$. To evaluate this expansion, it is convenient
to expand $\left|0_f\right>$ in terms of the
$\left|n_s\right>$. Representing $\varphi^2$ in terms of creation and
annihilation operators it is then a matter of straightforward, but
tedious algebra to evaluate the determinant. The result for short time
intervals is
\eqn\detsmall{
\det(-\partial_t^2+\omega^2(t))=
\exp\left[\int_{T_s}^{T_f}\!\!\omega
dt+{v^4T_s^2\over8\omega_s^4}\Delta T^2
+{v^4T_s\over24\omega_s^4}\left(-2\omega_s T_s+3-
6{v^2T_s^2\over\omega_s^2}\right)\Delta T^3 +{\cal O}(\Delta
T^4)\right].}
We have checked the consistency between this expansion and the WKB
expansion above in the region where $v^2T_s\Delta T\ll1$,
$\omega\Delta T\gg1$ (we checked all terms involving
$v^4$). From the determinant we calculate the short time expansion of
the phase shift,
\eqn\dmtshort{\eqalign{
\delta&=
\Delta T\left[{15\over16}{v^4\over r_s^7}+{315\over128}
{v^6\over r_s^{11}}+\ldots\right]
+\Delta T^2\left[\left(-{105\over32}{v^5\over
r_s^8}-{3465\over256}{v^7\over r_s^{12}}+\ldots\right)
(\hat{r}_s\cdot\hat{v})\right.\cr
&\left.+\left(-15{v^6\over r_s^{10}}-105{v^8\over r_s^{14}}
+\ldots\right)(\hat{r}_s\cdot\hat{v})^2\right]
+{\cal O}(\Delta T^3).\cr}}
We see that even at short time we reproduce the supergravity result to
leading order, the piece linear in $\Delta T$ is nothing but the
simple
adiabatic result from above, while we see deviations from this
approximation at order $\Delta T^2$. These deviations cannot be easily
written in form of a potential, which would have required $\delta$ to
take the form
\eqn\dalt{
\delta=-\int_{T_s}^{T_s+\Delta T}V_{\rm eff}(t)dt=-V_{\rm eff}(T_s)
\Delta T-{1\over2}\dot{V}_{\rm eff}(T_s)\Delta T^2+\ldots}
The new terms appearing at order $\Delta T^2$ are indicative of terms
in the effective Lagrangian that are not expressible as a single local
time integral.

If we consider the short time behavior for the case of vanishing
boundary conditions on the off-diagonal fields, we find a very
different result. We can evaluate the short-time determinant by
expanding $\chi(t)$ in a Taylor series, and solve \deteq\ order by
order in $\Delta T$. The first terms in the determinant will then be
proportional to $\omega^2$, $\omega^4$ and so on, and these terms all
cancel when multiplying the determinants for the phase shift. In fact
the phase shift vanishes all the way up to $\Delta T^8$ and we find
\eqn\dmtvvshort{
\delta_{vv}={v^4\over1575}\Delta T^8+{\cal O}(\Delta T^{10}).}
This strange result probably shows that this is a particularly unwise
choice of boundary conditions at short time intervals.

\newsec{Discussion}

When considering finite time amplitudes in Matrix theory, one must be
careful in specifying boundary conditions on the off-diagonal modes
that have no counterpart in supergravity. The most natural choice
seems to be to demand these modes to be in their ground state. In this
case we derive an effective potential between the gravitons that
differs from the one obtained from the infinite time calculations. The
leading order term, which is the one reproduced by supergravity, still
matches perfectly, providing a functional agreement between
supergravity and Matrix theory.
In comparing these Matrix theory and supergravity calculations, one
expects the low energy considerations to be valid only away from very
short time intervals. We find that for long, but finite time
intervals, there is a correction to the effective potential, depending
on the angle between $\vec{r}$ and $\vec{v}$,
compared to the one read off from the infinite time phase shifts.
There
are also new terms that are not expressible in terms of a potential,
whose role is not so clear.
In
particular the first subleading term in the potential, of order
$v^6/r^{11}$, is non-zero. Note that this term dominates the
$v^6/r^{14}$-term (which has been matched between two loop Matrix
theory and supergravity) at long distances.  Since supergravity should
be corrected by M-theory effects, it would be interesting to interpret
the term we found as such an M-theory correction.

This type of calculation should be straightforward to extend to
include spin effects and other objects than gravitons. We expect the
same features to show up also then, namely that the leading term
matches supergravity as in the infinite time case, with corrections to
the subleading terms. It would also be interesting to consider
extensions
to higher loop calculations.

\bigskip

\centerline{\bf Acknowledgements}

We are grateful to C.\ Callan, A.\ Guijosa, M.\ Krogh, S.\ Lee, and
S.\ Minwalla for useful discussions. This work was supported in part
by NSF grant PHY96-00258  and the Research Council of Norway.

\listrefs

\end